\documentclass[12pt]{article}
\renewcommand{\theequation}{\thesection.\arabic{equation}}
\def\annexe#1#2{\def\thesection{\Alph{section}}\section*{#2}
                \setcounter{section}{#1}  }

\begin{document}

\title{Noncommutative Planar Particle Dynamics
\\ with Gauge Interactions}

\author{J. Lukierski\\
Institute for Theoretical Physics, University of Wroc\l aw, \\
 pl. Maxa Borna 9, 50-204 Wroc\l aw, Poland
 \\
  email: lukier@ift.uni.wroc.pl
  \\ \\
P.C. Stichel\\
An der Krebskuhle 21\\ D-33619 Bielefeld, Germany \\
e-mail:pstichel@gmx.de
\\ \\
W.J. Zakrzewski
\\
Department of Mathematical Sciences,
 Science Laboratories,            \\
University of Durham, South Road,
Durham DH1 3LE, UK \\
 e-mail: W.J.Zakrzewski@durham.ac.uk
}

\date{\today}

\maketitle

\begin{abstract}
We consider two ways of introducing minimal Abelian gauge
interactions into the model presented in [1]. These two approaches are
different only if
 the second central charge of the planar Galilei group is nonzero.
  One 
  way
  leads to the standard gauge transformations and the other one to a generalized gauge theory with gauge
  transformations accompanied by time-dependent area-preserving
coordinate
  transformations. Both approaches, however, are related to each other by a classical
  Seiberg-Witten map 
  supplemented by a noncanonical transformation of
  the phase space variables for planar
particles. We also formulate the two-body problem in the model
with our generalized gauge symmetry and consider the case with both
CS and background electromagnetic fields, as it is used in the
description of fractional quantum Hall effect.

\end{abstract}

\section{Introduction}

Recently there has been a lot of interest
in considering quantum-mechanical and field-theoretic
models with noncommutative
space-time coordinates:
\begin{equation}\label{luz1.1}
 [ \widehat{x}_\mu, \widehat{x}_\nu ]
 = i\theta_{\mu\nu} (\widehat{x}) =
 i ( \theta^{(0)}_{\mu\nu} + \theta^{(1)}_{\mu\nu}
  {}^{\rho}\widehat{x}_\rho + \ldots )
\end{equation}

If $\partial_\rho \theta_{\mu\nu} (\hat{x}) \neq 0$ the
Poincar\'{e}
 symmetries
 with commutative translations do not preserve the relation
(\ref{luz1.1})
 and so the only case invariant under classical translations
$\hat{x}^\prime_\mu
  = \hat{x}_\mu + a_\mu$ ($a_\mu$ - $c$-numbers) is provided by
  $\theta_{\mu\nu} (\hat{x}) = \theta^{(0)}_{\mu\nu}$.
   Such a deformation, first introduced
   on the grounds of quantum gravity by
   Doplicher, Fredenhagen and Roberts [2], was
   further justified in $D=10$ string-theory moving in the background
   with a
  nonvanishing tensor field $B_{\mu\nu}$ [3,4]. However, it
is easy to see
  that even for constant value of the commutator (\ref{luz1.1})
   the noncommutativity of space-time breaks Lorentz
  invariance, i.e. $\theta^{(0)}_{\mu\nu}$
  is a constant tensor. If we assume
  that the relation (\ref{luz1.1}) is valid in all classical Poincar\'{e}
frames then this
  constant tensor should be described by a scalar parameter. The
  following two cases can be considered:

  i) $D=2$ relativistic theory, with classical Poincar\'{e} symmetries.
  In such a case
\begin{equation}\label{luz1.2}
  \theta^{(0)}_{\mu\nu} = \hbar \theta\, \varepsilon_{\mu\nu}\, ,
\end{equation}
  where $\varepsilon_{\mu\nu}$ is a $D=2$ covariant antisymmetric
tensor.

  ii) $D=2+1$  nonrelativistic theory, with a classical time variable
   and relations (1.1) applied to the $D=2$ space coordinates
   $x_i $ ($i=1,2$). In this case one gets
\begin{equation}\label{luz1.3}
  \theta_{ij} = \hbar \theta \, \varepsilon_{ij}\, .
\end{equation}
   It is known  that in a nonrelativistic Galilean-invariant theory the
space-time
   coordinates can be related to the Galilean boosts by the following
   relation [5]
\begin{equation}\label{luz1.4}
  {K}_i = m \,X_i^L\, .
\end{equation}
   The formulae (\ref{luz1.3}--\ref{luz1.4}) in a $D=2+1$ nonrelativistic
theory imply
   that the Galilean symmetry is endowed with two central charges:
    one standard describing mass $m$, and  the second ``exotic'', described
by
    the parameter $\theta$ in (\ref{luz1.3}). Moreover,  if we consider
the
    $(2+1)$--dimensional nonrelativistic  $c \to \infty$ limit of
   a $(2+1)$--dimensional relativistic theory, the parameter $\theta$
determines
     the value of the nonrelativistic Abelian $D=2$ spin \cite{luz6x}.

     The noncommutativity of position coordinates can be obtained as a
     consequence of canonical quantization of dynamical models.  Such
      a result is valid for string--inspired noncommutativity
      and for the $(2+1)$--dimensional Galilean models
      with noncommutative
       spatial coordinates. In our previous paper  [1] we have
      shown that a nonvanishing value of $\theta$ (see (\ref{luz1.3})) can
be
        introduced
      by the  following extension of the free classical $D=2+1$ particle
action
       ($\dot{a} \equiv {d \over dt } a$):

\begin{equation}\label{luz1.5}
  L = \frac{{m}\dot{x}^2_i}{2} - k \varepsilon_{ij}
  \dot{x}_{i} \ddot{x}_j\, .
\end{equation}

       The action (\ref{luz1.5}) contains higher derivatives and their
presence leads,
       after canonical quantization, to the introduction of
noncommutative position
        variables.

By comparison with formula (\ref{luz1.3}), one can show
that
\renewcommand{\theequation}{\thesection.5\alph{equation}}
\setcounter{equation}{0}
\begin{equation}\label{luz1.5a}
  {k}= - \frac{ \theta  m^2}{2}.
\end{equation}

\renewcommand{\theequation}{\thesection.\arabic{equation}}
\setcounter{equation}{5}

 The action
       (\ref{luz1.5}), in the Hamiltonian approach, is characterized by
  a    six-dimensional phase space with two canonical momenta
\renewcommand{\theequation}{1.6\alph{equation}}
\setcounter{equation}{0}
\begin{eqnarray}\label{luz1.6a}
p_i & = & \frac{\partial L}{\partial \dot{x}_i} -
\frac{d}{dt}\frac{\partial L}{\partial \ddot{x}_j} = m \dot{x}_i -
2 k\varepsilon_{ij} \ddot{x}_j
\\
\label{luz1.6b}
 \widetilde{p}_{i} & = & \frac{\partial L}{\partial
\ddot{x}_i} = k \epsilon_{ij} \dot{x}_j
\end{eqnarray}
\renewcommand{\theequation}{\thesection.\arabic{equation}}
\setcounter{equation}{6}
 which leads to the Hamiltonian
\begin{equation}\label{luz1.7}
  H = - \frac{m}{2k^2} (\widetilde{p}_j)^2 +\frac{1}{k}
  \widetilde{p}_{k} \epsilon_{kl} p_{l}\, .
\end{equation}

Introducing the variables
\begin{equation}\label{luz1.8}
  X^L_i = x_i - \frac{2}{m} \widetilde{p}_i,
  \qquad
  P_i = p_i, \qquad
  \widetilde{P}_i = \frac{k}{m} p_i +\epsilon_{ij}
  \widetilde{p}_j\,
\end{equation}
we get
\begin{equation}\label{luz1.9}
  H  = \frac{\overrightarrow{P}^2}{2 m}
  - \frac{m \overrightarrow{\tilde P}^2}{2k^2}
\end{equation}
and, considering  (\ref{luz1.6b}) as a constraint, we see that we
get the
following symplectic structure [1]:
\begin{equation}\label{luz1.10}
  \{ Y_A, Y_B \} =  \Omega_{AB}\, ,
\end{equation}
where
\begin{equation}\label{luz1.11}
\Omega = \pmatrix{
  \frac{2k}{m^2}\varepsilon & 1_2 & 0 \cr
  - 1_2 & 0 & 0 \cr
  0 & 0 & \frac{k}{2}\varepsilon
}
\end{equation}
and where $Y_A=\{X\sp{L}_i,P_k,\tilde P_l\}$.

We see that

i) the parameter $k$ introduces noncommutativity in the
coordinate sector\footnote{(1.10-11) describes classical Poisson 
brackets which are nonvanishing in the coordinate sector. For convenience,
we will refect to this fact here and in the rest of this paper as 
`noncommutativity' both in the quantum and in the classical case.}

ii) the dynamics splits into the decoupled sum of the dynamics in the 
physical sector ($X_i^L , P_i $ variables) and in the auxiliary sector
($\widetilde{P}_i$ variable).

In this paper we  consider the model (\ref{luz1.5}) with
electromagnetic interaction. Following the method of Faddeev and
Jackiw \cite{luz7x,luz8x} we rewrite the Lagrangian
(\ref{luz1.5}) in the first-order form, and introduce noncommutative
coordinates
\begin{equation}\label{luz1.12}
  X_i = X^L_i + \frac{2k}{m^2} \varepsilon_{ij} P_j\, ,
\end{equation}
which were recently introduced by Horvathy and Plyushchay
\cite{luz9x}. The noncommutative coordinates (\ref{luz1.12})
satisfy the relations (see (\ref{luz1.5a}))
\begin{equation}\label{luz1.13}
\left\{ X_i , X_j \right\} = - \frac{2k}{m^2} \varepsilon_{ij} = \theta
\varepsilon_{ij}
\end{equation}
and transform with respect to the Galilean boosts as components
of a Galilean two--vector.

 The electromagnetic interaction with a
magnetic potential can be introduced in two different ways:

i) By adding to the Lagrangian the term
\begin{equation}\label{luz1.14}
  L^{\rm int} = e A_i (X_i , t) \dot{X}_i\, .
\end{equation}
Such a way of introducing  electromagnetic
interaction can be interpreted as corresponding to the
modification of the  symplectic form of
the system which determines  the noncommutative phase-space
geometry (\ref{luz1.10}-\ref{luz1.11}) \cite{luz10x}.

ii) One can introduce the minimal EM coupling by
 the replacement
\begin{equation}\label{luz1.15}
  H_0 = \frac{P\sp2}{2m} \to \frac{\vec{\cal P}^2}{2m} =
  \frac{1}{2m} \left( \overrightarrow{P} - e\overrightarrow{A}
  (X_i , t) \right)^2
\end{equation}
and preserve the symplectic structure
(\ref{luz1.10}-1.13). In such a way  the interaction
 does not modify the noncommutative geometry, but changes Abelian gauge
transformations.

The main aim of this paper is to consider the case ii), which is related to
 models describing the quantum Hall effect, with generalised gauge
transformations accompanied by area - preserving transformations
(see e.g. \cite{luz11x} -- \cite{luz13x})\footnote{Area-preserving transformations are the symmetry transformations for electrons in the lowest Landau level. They have been introduced in [14] and recently studied in [15].}
After considering in Sect.
2 the first order formalism for our model from [1]
 and the canonical structure of both models, i) and ii),  we introduce  the area reparametrization -
invariant formalism. In
Sect. 3 we show that both possibilities are related to each other
by a classical Seiberg-Witten (SW) map [3] 
    supplemented by a noncanonical transformation of phase space variables
    for planar particles. In such a way we
recover the known
definition of covariantized coordinates [16] 
 describing the coordinate part of
 the noncanonical transformation in the
phase space describing planar particles. In Sect 4 we  consider
the Chern-Simons (CS) gauge interactions of planar particles and
formulate the dynamics of the corresponding two-body problem. This leads to the
 deformed anyonic dynamics which might then be applied to the description
of the quantum Hall effect. In Sect 5 we consider our model with
statistical CS fields in the electromagnetic background. We note
 that for the  critical value of the magnetic background field
strength we obtain the description of lowest Landau level for
Quantum Hall Effect. In the last section we comment on the second
quantization of our model [1] and outline the relativistic
generalization to $D= 3+1$. Finally, in an appendix we introduce a gauge 
field-dependent dreibein formalism.

\section{Two Ways of Introducing Minimal Gauge Couplings}
\setcounter{equation}{0}

 Following Faddeev-Jackiw's  method of
describing Lagrangians with higher order derivatives [8] we
describe, equivalently, the action (\ref{luz1.5}) as (see
[1])\footnote{For simplicity we give for all the particles the
same mass ($m=1$ in appropriate units) and use $\theta$
defined by (\ref{luz1.5a}) ($\theta = - 2k$) instead of $k$ as
the second central charge.}
\begin{eqnarray}\label{luz2.1}
  L^{(0)} &= & P_i (\dot{x}_i - y_i) + \frac{y_i^2}{2} + \frac{\theta}{2}
  \varepsilon_{ij} y_i \dot{y}_j =
\cr & = & P_i \dot{x}_i  + \frac{\theta}{2} \varepsilon_{ij} y_i
\dot{y}_j - H(y,P)\, ,
\end{eqnarray}
where
\begin{equation}\label{luz2.2}
H(y,P) = - \frac{y^2}{2} + P_i y_i\, .
\end{equation}
Using the variables [9]
\begin{eqnarray}\label{luz2.3}
Q_i & = & \theta (y_i - p_i ) \cr X_i & = & x_i +\varepsilon_{ij}
Q_j\, ,
\end{eqnarray}
we see that our Lagrangian separates into two disconnected
parts describing the ``external" and ``internal" degrees of freedom.
Thus we have
\begin{equation}\label{luz2.4}
  L^{(0)} = L_{\rm ext}^{(0)} + L_{\rm int}^{(o)}\, ,
\end{equation}
with
\renewcommand{\theequation}{\thesection.5\alph{equation}}
\setcounter{equation}{0}
\begin{eqnarray}\label{luz2.5a}
L^{(0)}_{\rm ext} & = & P_i \dot{X}_i  + \frac{\theta}{2}
\varepsilon_{ij} P_i
\dot{P}_j - H^{(0)}_{\rm ext} \\
 L^{(0)}_{\rm int}
 & = &  \frac{1}{2\theta}\varepsilon_{ij}Q_i \dot{Q}_j -H^{(0)}_{\rm
int}
\end{eqnarray}
\renewcommand{\theequation}{\thesection.\arabic{equation}}
\setcounter{equation}{5} where
\begin{equation}\label{luz2.6}
H^{(0)}_{\rm ext} = \frac{1}{2} \overrightarrow{P}^2\, , \qquad
H^{(0)}_{\rm int} = - \frac{1}{2\theta^2} \overrightarrow{Q}^2\, .
\end{equation}

From (\ref{luz2.5a}-\ref{luz2.6}) we obtain the following Poisson
brackets (PBs) of the independent sets of external and internal phase space
variables:
\begin{eqnarray}\label{luz2.7}
\{X_i, X_j\} & = & \theta \varepsilon_{ij}\, , \cr
 \{X_i, P_j\} & = & \delta_{ij}\, ,\cr
  \{P_i, P_j\} & = & 0     \, ,
\end{eqnarray}
and
\begin{equation}\label{luz2.8}
  \{Q_i, Q_j \} = - \theta \, \varepsilon_{ij}\, ,
\end{equation}
with all other PBs vanishing.

Having separated off the ``internal" degrees of freedom (i.e. $L_{int}$) we now proceed to couple in the electromagnetic field. We couple it to the
``external" sector only. Hence in the remainder of this paper we shall
not be concerned with the ``internal" sector of the theory (described
by $Q_i$ and $L\sp{(0)}_{int}$). We note 
first that the action (\ref{luz2.5a}) describes the
model by Duval and Horvathy [10], with the symplectic structure
given by the following Liouville form
\begin{equation}\label{luz2.9}
  \Omega = P_i dX_i + \frac{\theta }{2}\varepsilon_{ij}P_i dP_j
  - H_{\rm ext}^{(0)} dt\, .
\end{equation}

The minimal coupling to the gauge  field $A_\mu(
\overrightarrow{x},t) = (A_i ( \overrightarrow{x},t), A_0(
\overrightarrow{x},t))$ can  be introduced in the following two
ways:

\subsection{ Duval-Horvathy model}

One replaces the one-form (\ref{luz2.9}) by:
\begin{equation}\label{luz2.10}
  \Omega \to \Omega_e = \Omega +
  e(A_i dX_i +A_0 dt)\, ,
\end{equation}
which corresponds to the addition of (\ref{luz1.14}). Introducing
$dX_\mu = (dX_i, dt)$ the modification (\ref{luz2.10}) leads to
the symplectic form with a standard addition corresponding to the
minimal EM coupling
\begin{eqnarray}\label{luz2.11}
  \omega & = & d \Omega = dP_i \wedge dX_i + \frac{\theta}{2}\varepsilon_{ij}
   dP_i
   \wedge dP_j - dH_{\rm ext}^{(0)} dt
   \cr
&& +  e (\frac{1}{2} F_{ij} dX_i
 \wedge dX_j - E_i dX_i \wedge dt
)
  \, ,
\end{eqnarray}
where
\begin{equation}\label{luz2.12}
  F_{ij} = \partial_i A_j - \partial_j A_i = \varepsilon_{ij} B\,
  ,
  \qquad
  E_i = \partial_i  A_0 -
   \partial_t A_i \, .
\end{equation}

It is easy  to see that the symplectic form (\ref{luz2.11}) is
invariant under standard gauge transformations
\begin{equation}\label{luz2.13}
  A_i \to A^\prime_i = A_i + \partial_i \Lambda \, ,
  \qquad
A_0 \to A^\prime_0 = A_0 + \partial_t \Lambda \, .
\end{equation}

The Lagrangian corresponding to (\ref{luz2.10}) now becomes

\begin{equation}\label{luz2.14}
  L_{\rm ext} = L_{\rm DH} = (P_i + eA_i) \dot{X}_i +
  \frac{\theta}{2}\varepsilon_{ij} P_i \dot{P}_j -
  \frac{1}{2} \overrightarrow{P}^2 + eA_0 \ ,
\end{equation}
which may be brought by the point transformation
\begin{equation}\label{luz2.15}
  P_i \to { P}_i' = P_i + eA_i \, ,
\end{equation}
to the equivalent form:
\begin{equation}\label{luz2.16a}
  L_{\rm DH} =  { P}_i' \dot{X}_i +  \frac{\theta}{2}
\varepsilon_{ij}
  ({ P}_i' - eA_i)( \dot{ P}_j' - e \frac{d}{dt} A_j) + eA_0
  - \frac{1}{2}( { P}_i' - eA_i)^2.
\end{equation}

 The Lagrangian (\ref{luz2.14}) is quasi-invariant under standard
local gauge transformations (\ref{luz2.13}):
\begin{equation}\label{luz2.16}
  L_{\rm ext} \to L_{\rm ext}^{\prime} = L_{\rm ext} +
  \partial_i \Lambda \dot{X}_i + \partial_t \Lambda =
  L_{\rm ext} +\frac{d}{dt} \Lambda \, .
\end{equation}
The modification (\ref{luz2.10}), (\ref{luz2.11}) has been
considered in [10] and it leads to the modification of the PB
structure (2.7) [10]:
\begin{eqnarray}\label{luz2.17}
&  \{X_i, X_j\} & =  \  \frac{\theta \varepsilon_{ij}}{1 - e\theta
B}\, , \cr
& \cr
 & \{X_i, P_j\} & = \  \frac{\delta_{ij}}{1 - e \theta B}\, ,
\cr
& \cr
  & \{P_i, P_j\} & = \  \frac{e B\varepsilon_{ij}}{1 - e \theta B}\, .
\end{eqnarray}

\subsection{Model with generalized gauge transformations}

The other possibility of a minimal coupling follows from the assumption that
the symplectic structure 
(\ref{luz2.7}) remains unchanged. This is the case if we insert
 the minimal substitution\footnote{The gauge fields in this model  we shall denote by hat $(\hat{A}_\mu, \hat{F}_{n\nu})$ in order to distinguish them from the corresponding quantities in the model of Duval and Horvathy [10].}
\begin{equation}
P_i \to {\cal P}_i\,=\,P_i - e\hat{A}_i
\end{equation}
$$ H_{\rm ext}^{(0)} \to H_{\rm ext}^{(0)} -e \hat A_0$$
into  the
free Hamiltonian $H_{\rm ext}^{(0)}$ only.

In this way we get, in place of (\ref{luz2.16a}), the following Lagrangian
\begin{equation}\label{luz2.18}
  \widetilde{L}_{\rm ext} = P_i \dot{X}_i + \frac{\theta}{2}
  \varepsilon_{ij}   P_i \dot{P}_j - \frac{1}{2} (P_i - e\hat{A}_i)^2 +
  e\hat{A}_0 \, .
\end{equation}

The difference between both Lagrangians is in the 2nd term of (2.20). $L_{\rm DH}$ (2.16) arises from $L\sp{(0)}_{ext}$ by performing
the minimal substitution (2.19) not only in $ H_{\rm ext}^{(0)}$ but also
 in the
second term of $L\sp{(0)}_{ext}$.


We note that the symplectic structure described by
(\ref{luz2.7}) is invariant under the following 
infinitesimal time-dependent area - preserving
- local coordinate transformations
\begin{equation}\label{luz2.19}
  \delta X_i = - e\theta \varepsilon_{ij} \partial_j
   \Lambda (\overrightarrow{X},t) \qquad
  \delta P_i = e \partial_i  \Lambda (\overrightarrow{X},t),
\end{equation}
where $\Lambda$ is infinitesimal.

If we supplement (2.21) by the transformation
of the gauge fields
\begin{eqnarray}\label{luz2.20}
  \delta \hat{A}_\mu
    (\overrightarrow{X},t): &= & \hat{A}_\mu^\prime
      (\overrightarrow{X} + \delta \overrightarrow{X},t) - \hat{A}_\mu
        (\overrightarrow{X},t)
\cr
    &      = & \partial_\mu \Lambda (\overrightarrow{X},t)  \, ,
\end{eqnarray}
it is easy to check that the Lagrangian (\ref{luz2.18}) is
quasi-invariant
\begin{equation}\label{luz2.21}
  \delta \widetilde{L}_{\rm ext} = e \frac{d}{dt}
  ( \Lambda +\frac{\theta}{2} \varepsilon_{ij}
  \partial_i \Lambda P_j ) \, .
\end{equation}

We note that (2.22) differs from the standard gauge transformation (2.13)
by the simultaneous coordinate
transformation\footnote{For the mixing of gauge and coordinate transformation see also Jackiw et al. [17]} (2.21). For the corresponding change
$\delta_0\hat A_{\mu}$ of the gauge field at fixed $\vec X$ we obtain from (2.21-22)
\begin{eqnarray}
 \delta_0\hat A_{\mu}(\vec X,t):= \hat A_{\mu}'(\vec X,t)-
 A_{\mu}(\vec X,t)\cr
=\partial_{\mu}\Lambda(\vec X,t)+e\{ A_{\mu}(\vec X,t),\Lambda(\vec X,t)\}
\end{eqnarray}
in place of (2.13). Therefore we call the transformation (2.24) a generalised
gauge transformation. In deriving (2.24) from (2.22) we have used the PBs (cp. (2.7))
\begin{equation}
\{g,f\}:\,=\,\theta \,\epsilon_{ij}\,\partial_ig\,\partial_jf
\end{equation}
for two generic functions $f$ and $g$.

The equations of motion (EOM) derived from (\ref{luz2.18}) are
given by
\begin{eqnarray}\label{luz2.22}
  \dot{X}_i & = & - \theta \varepsilon_{ij} [ e(P_k - e\hat{A}_k)
  \partial_j \hat{A}_k + e\partial_j \hat{A}_0] + P_i -e\hat{A}_i\, ,
  \cr
  \dot{P}_i & = & e(P_k - e\hat{A}_k) \partial_i \hat{A}_k
   + e\partial_i \hat{A}_0
  \, ,
\end{eqnarray}
which, having made use of (\ref{luz2.7}), can be  put into the Hamiltonian form
\begin{equation}\label{luz2.23}
  \dot{X}_i = \{ X_i, H\}\, , \qquad
  \dot{P}_i  = \{ P_i, H\}\, ,
\end{equation}
where
\begin{equation}\label{luz2.24}
  H= \frac{1}{2}(P_i - e\hat{A}_i)^2 - e\hat{A}_0\, .
\end{equation}


 Let us rewrite  the EOM (\ref{luz2.22}) in terms of our new
 variable ${\cal P}_i$ (2.19).

We obtain
\begin{equation}\label{luz3.1}
  \dot{{\cal P}}_i = e( \widehat{F}_{ik} {\cal P}_k + \widehat{F}_{i0}
  )\, ,
\end{equation}
with the invariant field strength\footnote{We draw attention to the difference
from the model
of Duval et al. [10] which has a standard Abelian field strength.}

\begin{equation}\label{luz3.2}
  \widehat{F}_{\mu\nu} : = \partial_\mu \widehat{A}_\nu -
  \partial_\nu \widehat{A}_\mu +
e \{\hat A_{\mu},\hat A_{\nu}\}
\end{equation}
and
\begin{equation}\label{luz3.3}
\dot{X}_i + e\theta \varepsilon_{ij}
\partial_j \widehat{A}_0
= {\cal P}_k (\delta_{ik}  - e\theta \varepsilon_{ij}\partial_j
A_k)\, .
\end{equation}

\section{Seiberg-Witten (SW) Map 
and the Equivalence of the Two Planar Particle Models with
Noncommutative Structure}

\setcounter{equation}{0}

 In this section we show that our
model, (\ref{luz2.18}), and the one of Duval et al., (\ref{luz2.14}),
are related to each other by a noncanonical transformation of the phase space variables $(X_i, P_i) \to (\eta_i, {\cal P}_i)$ supplemented
by a classical SW map between the corresponding gauge potentials.

Let us introduce, besides the invariant ${\cal P}_i$ given by formula (2.19), the invariant
particle coordinates as follows\footnote{cp. [16,19] for the case of noncommutative gauge theories} (cp [13]):
\begin{equation}\label{luz4.1}
  \eta_i (\overrightarrow{X},t):= X_i + e\theta \varepsilon_{ij}
  \widehat{A}_j (\overrightarrow{X},t)\, .
\end{equation}

Clearly from (2.21-22) we obtain
\begin{equation}
\delta \eta_i=0
\end{equation}
but at fixed $\vec X$ the fields $\eta_i$ transform as 
\begin{equation}
\delta \eta_i\,=\,e\{\eta_i,\Lambda\}.
\end{equation}

It is easy to check that the new phase-space variables ($\eta_i,
{\cal P}_i$) satisfy the noncanonical Poisson brackets
(2.18)

\begin{equation}\label{luz4.2}
\{ \eta_i, \eta_j\} = \frac{\theta \epsilon_{ij}}{1-e\theta B(\vec{\eta},t)}\ ,
\end{equation}
$$
\{ \eta_i, {\cal P}_j\} = \frac{\delta_{ij}}{1-e\theta B(\vec{\eta},t)}\ , \quad \{ {\cal P}_i, {\cal P}_j\} = \frac{e\epsilon_{ij} B(\vec{\eta},t)}{1-e\theta B(\vec{\eta},t)}
$$
with the field $B$ defined by (cp. [20])
\begin{equation}\label{luz4.3}
B(\vec{\eta},t) = \frac{\hat{B} (\vec{X},t)}{1+e\theta \hat{B}(\vec{X},t)}
\end{equation}
where $X_i$ is a function of $\eta_i$ as follows from (3.1).

The relations (3.4) as well as (2.7) describe, after quantization, two different quantum phase spaces with noncommutative position sectors.

With $(\eta_i, {\cal P}_i )$ as the new noncanonical phase-space
variables our $L$ (\ref{luz2.18}) becomes
\begin{equation}\label{luz4.2}
  L = \widehat{L}_{\rm part} + \frac{\theta}{2}
  \varepsilon_{ij}{\cal P}_i \dot{{\cal P}}_j \, ,
\end{equation}
where $\widehat{L}_{\rm part}$ is given by the $\theta$-deformed
particle Lagrangian in the presence of gauge fields defined in [13],
i.e.
\begin{equation}\label{luz4.3}
\widehat{L}_{\rm part} = {\cal P}_i \dot{\eta}_i - \frac{1}{2}
{\cal P}_i^{2} + e( \widehat{A}_i \dot{X}_i + \widehat{A}_0 +
\frac{e\theta}{2} \varepsilon_{ij} \widehat{A}_i  \frac{d}{dt}
\widehat{A}_j) - \frac{1}{2} \frac{d}{dt} (e \theta
\varepsilon_{ij} {\cal P}_i \widehat{A}_j )\, .
\end{equation}
Moreover,  we neglect the total time-derivative term which is
irrelevant for EOM.

In order to express $L$ in terms of $(\eta_i, {\cal P}_i)$ we have
to introduce a map
\begin{equation}\label{luz4.4}
  \widehat{A}_\mu ( \overrightarrow{x},t) \to
  A_\mu(\overrightarrow{\eta}, t)\, .
\end{equation}
In accordance with [13] we define (\ref{luz4.4}) by the requirement
\begin{equation}\label{luz4.5}
\widehat{A}_i \dot{X}_i + \widehat{A}_0 + \frac{\theta}{2}
\varepsilon_{ij} \widehat{A}_i \frac{d}{dt} \widehat{A}_j
= A_i ( \overrightarrow{\eta}, t) \dot{\eta}_i + A_0 (
\overrightarrow{\eta},t)\, .
\end{equation}
Eliminating at the l.h.s. of (\ref{luz4.5}) $\dot{X}_i$ in favour
of $\dot{\eta}_i$  we obtain, by comparing  the coefficients of
$\dot{\eta}_i$ as well as of unity at both sides of (\ref{luz4.5}),
the relations
\begin{equation}\label{luz4.6}
  A_k \left( \overrightarrow{ \eta} ( \overrightarrow{X},t),t \right)
  = \frac{1}{2}
  \hat{A}_l ( \overrightarrow{X},t)
  \left(\delta_{kl} + \frac{e_{kl}( \overrightarrow{X},t)}
  {1 + e\theta \widehat{B}}
  \right)
\end{equation}

\begin{equation}\label{luz4.7}
  A_0 \left( \overrightarrow{ \eta} ( \overrightarrow{X},t), t \right)
  = \hat{A}_0 ( \overrightarrow{X},t)
-   \frac{e \theta}{ 2(1+e\theta \widehat{B})}
   \widehat{A}_{l} ( \overrightarrow{X},t)
\varepsilon_{kj} \partial_t \widehat{A}_j ( \overrightarrow{X},t)
e_{kl} ( \overrightarrow{X},t)\, ,
\end{equation}
expressed in terms of the inverse dreibein, which we discuss in more detail in the Appendix (see (A.6) and
also [13], Eq. (24)).

From (3.10-11) we derive a simple relation between the corresponding field
strengths (cp. [20])
\begin{equation}\label{luz4.10}
F_{\mu\nu} (\vec{\eta},t)= \frac{\hat{F}_{\mu\nu} (\vec{X} ,t)}{1+e\theta \hat{B} (\vec{X},t)}\ .
\end{equation}

The relations (\ref{luz4.6}) and (\ref{luz4.7})  are just the
classical limits of an inverse SW-map defined by replacing in the SW
differential equation ([3], eq. (3.8)) star products by ordinary products (cp. ([21], sect. 2) and ([20], sect. 4.1)).  They give us  the
required relation between our Lagrangian given by (\ref{luz2.18})
and the one of Duval and Horvathy denoted by $L_{DH}$ and given by
(\ref{luz2.14})
\begin{equation}\label{luz4.8}
  L \left( \widehat{A}_\mu (\overrightarrow{X},t), \dot{\overrightarrow{X}},
  \overrightarrow{X}, \overrightarrow{P}, \dot{\overrightarrow{P}})\right)
  =L_{DH} (A_\mu ( \overrightarrow{\eta},t),
  \overrightarrow{\eta},
  \dot{\overrightarrow{\eta}}, \overrightarrow{{\cal P}},
  \dot{\overrightarrow{{\cal P}}})).    \,
\end{equation}

Thus we see that the relations (3.10) and (3.11),
supplemented by the transformation (3.1) and (2.19),
describe within a classical framework the SW map relating
the planar particle dynamics in the presence of Abelian gauge
fields in two different noncanonical phase spaces with two different
symplectic structures.
These symplectic structures are either gauge field independent
(cp. (2.7)) or gauge field dependent (cp. (3.4)), (cp. [20,21]).
A characterization of the SW map
as relating two different symplectic structures
has been considered also earlier (see e.g. [21,22])
and provides an extension of the original
formulation in terms of infinitesimal gauge transformations [3]
in the presence of particle coordinates.

The relation (\ref{luz4.8}) is the central result of our paper. We
see that the two models describing different possibilities of
introducing minimal
 electromagnetic interaction, one with the standard gauge
 transformations (see (\ref{luz2.13})) and the other one with the generalized
 gauge transformations (see (2.24)), may be transformed
 into each other by a local Seiberg-Witten
 transformation accompanied by a change of phase space variables
 in the particle sector.
 It should be stressed that if $\theta \neq 0$, in both phase spaces, the
 Poisson brackets in the coordinate sector imply  noncommutative space coordinates.
 In this way we 
 have achieved an extension to $\theta \neq 0$ of
 a classical SW map for standard point particles with commuting
 space coordinates %
 considered in
 [13].

The total action is obtained if we further add a pure gauge part of the action
(Maxwell, Chern-Simons etc.), with corresponding symplectic structures
(and, ultimately, one can add also our ``internal" Lagrangian $L\sp{(0)}_{int}$).
 In particular if the gauge field actions transform into each
 other by the SW-map (\ref{luz4.6}-\ref{luz4.7}),  the  particle
 trajectories with gauge
 interaction in the respective phase-spaces are classically equivalent i.e.
 may be expressed equivalently in two noncanonical phase space frameworks.
 It should be added
 that such a  classical  equivalence might become invalid
 after quantization due to the operator ordering problems providing
 $\theta$-dependent quantum corrections to the particle
 interactions.

It is worth noting that, using arguments similar to ours, Jackiw et al.
have
presented in a very recent paper [19] the Seiberg-Witten map relating
the Lagrange and Euler pictures in the presence of gauge fields 
for
another dynamical model: the field-theoretical formulation of fluid
mechanics.

\section{Chern-Simons Gauge Interaction and the Two-Body Problem}
\setcounter{equation}{0}

 In this Section we derive the
dynamics  for two identical particles described by
our model (\ref{luz2.18}) interacting via Chern-Simons (CS) gauge
interactions.

Let us start with the CS-action of a $\widehat{A}_\mu$ field
invariant with respect to the generalized gauge transformation
(2.24).

We have (cp.        [11],    [13])
\begin{equation}\label{luz5.1}
  L_{\rm CS} = \frac{\kappa}{2} \int d^2 x        \,
  \varepsilon^{\mu\nu\rho} \widehat{A}_\mu \left(
  \partial_\nu \widehat{A}_\rho +
  \frac{e }{3} 
\{\hat A_{\nu},\hat A_{\rho}\}
 \right) \, .
\end{equation}

The extra (unusual) term in this expression is required by our
generalised gauge invariance as discussed in [11] and [13]. Its origin 
can be traced to the appearance of an extra term in (2.30).

Next we consider the following total Lagrangian
\begin{equation}\label{luz5.2}
  L_{\rm tot} = \sum\limits_{\alpha =1}^{2} \widetilde{L}_{{\rm
  ext}, \alpha} +  L_{\rm CS} \, 
\end{equation}
with each of $\widetilde{L}_{\rm  ext}$ given by (2.20).

The variation of $L_{\rm CS}$ with respect to the Lagrange multiplier field
$\widehat{A}_0$ leads to the well known Gauss constraint
\begin{equation}\label{luz5.3}
  \epsilon_{ij}\widehat{B}(\overrightarrow{x},t) = 
\hat F_{ij}= - \epsilon_{ij}\frac{e}{\kappa} \,
  \sum\limits_{\alpha = 1}^{2} \delta( \overrightarrow{x} -
  \overrightarrow{X}_{\alpha} )\, .
\end{equation}

Modulo asymptotic parts, which do not contribute to the
Hamiltonian describing relative particle motion, we obtain a
solution of (\ref{luz5.3}) for $\widehat{A}_k$ at the particle
position $ \overrightarrow{x} = \overrightarrow{X}_{1\atop 2}$ in the form [13]
\begin{equation}\label{luz5.4}
  \widehat{A}_{k}(X_{1\atop 2}) = \pm \varepsilon_{kj} (X_1 - X_2)_j
  \chi (|\overrightarrow{X}_1 - \overrightarrow{X}_2 |)\, ,
\end{equation}
with
\begin{equation}\label{luz5.5}
\chi(R) = \frac{1}{e \theta }
 \left( 1 - \left( 1 -
\frac{\widetilde{\theta}}{R^2} \right)^{1/2} \right)= \frac
{1}{2} \frac{e}{\pi \kappa} \frac{1}{R^2} \left( 1+ \frac
{1}{4} \tilde{\theta} \frac{1}{R^2} + 0(\theta^2)
\right) \, ,
\end{equation}
where $R=|\overrightarrow{X}|$ and
$${\widetilde{\theta}} : = \frac{e^2 \theta}{\pi \kappa}.
$$

With (\ref{luz5.4}-\ref{luz5.5}) and the position and momentum
variables for the relative motion
\begin{equation}\label{luz5.6}
  \overrightarrow{X}:= \overrightarrow{X}_1 - \overrightarrow{X}_2\, ,
\qquad
  \overrightarrow{P}:= \frac{1}{2} \left( \overrightarrow{P}_1
    - \overrightarrow{P}_2 \right)\, ,
\end{equation} and
by applying the Legendre transformation to (\ref{luz5.2}) and
using the Gauss - constraint (\ref{luz5.3}) we obtain the following
Hamiltonian for the relative motion
\begin{equation}\label{luz5.7}
  H = \overrightarrow{P}^2 + 2e \left( \varepsilon_{ij} X_i P_j +
  \frac{R^2}{\theta}\right) \chi (X_k)
  - \frac{e^2}{\pi \kappa \theta}
\end{equation}
$$ = \overrightarrow{P}^2 + \frac{e^2}{\pi \kappa}
 \varepsilon_{ij} X_i P_j \frac{1}{R^2} + \frac{e^4}{4\pi^2 \kappa^2 R^2}
  + O \left( \theta \right) ,$$
i.e. in the leading order of the $\theta$-expansion we reproduce the known anyonic Hamiltonian.

The phase-space variables for the relative motion (\ref{luz5.6}) obey,
according to (\ref{luz2.7}), the Poisson bracket relations
\begin{eqnarray}\label{luz5.8}
&  \{X_i, X_j\} &=  \ 2 \theta \varepsilon_{ij}\, , \cr
 & \{X_i, P_j\} & = \  \delta_{ij}\, ,
\cr &  \{P_i, P_j\} & =  \  0 \, .
\end{eqnarray}

In order to quantize the Hamiltonian system (4.7-8) we proceed in three
steps:

i) We replace the classical structure (4.8) by commutators of the
corresponding operators
\begin{equation}
\{A,\,B\}\quad \rightarrow \quad {1\over i\hbar}[\hat A,\,\hat B],
\end{equation}
where $\hat A$, $\hat B$ denote the quantized variables.\footnote{We hope that there is no confusion here with the hat introduced before - for the field
quantities of our model}

ii) We solve the ordering problem arising from the noncommuting position and momentum variables by symmetrization
\begin{equation}
P_i\chi(X_k)\,\rightarrow \,{1\over 2}\left(\hat P_i\hat \chi(\hat X_k)\,+\,\hat \chi(\hat X_k)\hat P_i\right).
\end{equation}

iii) We replace the operator-valued functions $\hat f(\hat X_k)$,
$\hat g(\hat X_k)$ of noncommuting position variables $\hat X_k$
with local multiplication by functions $f(y_k)$, $g(y_k)$
depending on commuting position variables $y_k$ and the nonlocal
Moyal-star product
\begin{eqnarray}
& \hat f(\hat X_k) \hat g(\hat X_k) \, \longleftrightarrow \,
f(y_k)* g(y_k) : =\cr 
& \cr
& f(y_k)\exp \left( i\hbar \theta \epsilon_{ij}
\overleftarrow{\partial}_i \overrightarrow{\partial}_j \right)
g(y_k) = : f\left( y_k - \theta \varepsilon_{kl} \hat P_l
\right) g(y_k),
\end{eqnarray}
where
\begin{equation}\label{luz6.8}
\hat P_i := \,{\hbar\over i}\partial_i,
\end{equation}
with $\partial_i:={\partial \over \partial y_i}$.

Such a quantization procedure leads to the Schr\"odinger equation
\begin{equation}
(-\hbar\sp2\Delta \,-\,{e\sp2\over \pi k\theta}\,-\,E)\psi
\,+\,2e\epsilon_{ij}(y_i\chi(y))*\hat P_j\psi\,+\,{2e\over \theta}(y\sp2\chi(y))*\psi\,=\,0.
\end{equation}

In deriving (4.13) we have used the property that $\chi$ is a
function of only $y:=\vert \vec y\vert$ (see (4.5)) and thus
\begin {equation}
\epsilon_{ij}\,y_i\,(\hat P_j\chi)\,=\,0.
\end{equation}

In this Section we have been considering the gauge interaction
between two identical particles, with the same charge $e$. An
interesting question now arises, as to whether the Poisson
bracket (4.8) for relative coordinates should depend on the choice
$e_1, e_2$ of charges at the points $\vec{X}_1, \vec{X}_2$. If we
observe that $\theta$ is geometrically similar to the mass
parameter, which is also a Galilean central charge, one can
assume, by analogy, that $\theta$ differs for particles with
different electric charges. In order to obtain for $N$ planar
particles the invariant action (4.2) we are led to the consistant
replacement $\theta \to \frac{\theta}{e}$ in the formulae of
Sect. 2-4. In such a case
one gets for relative coordinates (4.6) in the $N=2$ case %
the following modification of the  first formula (4.8)
\begin{equation}
\{ X_1, X_i\} = \theta \left( \frac{1}{e_1} + \frac{1}{e_2}\right)\ ,
\end{equation}
i.e. if $e_1 = - e_2$ we obtain $\{ X_1, X_2\} = 0$,
 in agreement with the conclusions of [23].

\section{Application: Statistical planar CS  gauge action
and external electromagnetic background fields}
\setcounter{equation}{0}

\subsection{Physical background}

It is known that CS gauge transformations as well as CS gauge fields in the
$D=2+1$ Hamiltonian framework are used for the description of the Fractional Quantum Hall Effect (FQHE) (see e.g. [24,25]) and represent flux tubes attached to electrons forming basic fermionic quasiparticles - composite fermions (CF). However, formally such CS gauge fields are gradients, i.e. pure gauge, the gauge functions are multivalued and from the Stokes theorem it follows that the CS gauge field strength is nonzero.
In what follows these gauge fields $A_\mu^{CS}$, which dress the electrons
in the Hamiltonian formulation of FQHE, will be called statistical CS fields.

In a general case one can embedd the system of CFs in an
 external  electromagnetic background field $A_\mu^{ext} (X)$, i.e. add to the
 CS actions considered in sect. 2 additional gauge field couplings.
 One can proceed in two ways:

i) By modifying the minimal substitution (2.19) in the Hamiltonian
\begin{equation}
P_i \to P_i - e\hat{A}_i\ \  {\bf{\longrightarrow}}\ \  P_i \to  P_i - e\hat{A}^{tot}_i\ ,
\end{equation}
where $\hat A_i\sp{tot}$ turns out to be a nonlinear function of $\hat A_i\sp{CS}$ and $\hat A_i\sp{ext}$ as given below.

ii) By adding to the Lagrangian (2.20) the background field term in the form of 
(1.14).

We shall consider below these two couplings in our model, (2.20),
which is invariant with respect to the area-preserving coordinate transformations (2.21-22).

\subsection{Minimal coupling (5.1)}

Our main point here is, that for such a coupling, the gauge 
fields in our model 
(for $\theta \not= 0$) are nonadditive. 

Firstly, let us observe
that in the DH Lagrangian (2.14) the gauge fields are coupled
linearly, i.e. one gets Abelian addition formula
\begin{equation}
A^{tot}_\mu = A^{CS}_\mu + A^{ext}_\mu
\end{equation}
but the gauge fields $\hat{A}^{tot}_\mu$ in our model will be the solution of the relations (3.10-11) and so (see (3.13))
\begin{equation}
L(\hat{A}^{tot}_\mu (\vec{X},t), \vec{X}, \dot{\vec{X}},  \vec{P},
\dot{\vec{P}} ) = L_{DH} (A_\mu^{CS} (\vec{\eta},t) + A_\mu^{ext}
(\vec{\eta},t)\ , \vec{\eta}, \dot{\vec{\eta}}, \vec{\cal{P}},
\dot{\vec{\cal{P}}})
\end{equation}

In order to have insight into the nonlinear structure of our
decomposition of $\hat A_{\mu}\sp{tot}$ we determine the SW map (3.10-11)
for
$\hat A_{\mu}\sp{tot}$ in the lowest order of the $\theta$ expansion using
(3.1) (cp. [3]):
\begin{equation}
\hat A_{\mu}\sp{tot}(\vec x,t)\, =\,  A_{\mu}\sp{tot}(\vec x,t)
\,-\,{e\theta\over 2}\epsilon_{ik} \, A_{i}\sp{tot}
(\partial_k  A_{\mu}\sp{tot}\, +\, F_{k\mu}\sp{tot})\, +\, O(\theta\sp2),
\end{equation}
where
$A_{\mu}\sp{tot}$ is given by (5.2) and the field strength $F_{k\mu}\sp{tot}$
is related to $A_{\mu}\sp{tot}$ by (2.12).

The analogue of (5.4) for the field strength has been given in
a closed form in (3.12), i.e. we have
\begin{eqnarray}
\hat B\sp{tot}(\vec X,t) \, =\, & {B\sp{tot}(\vec \eta, t) \over 1 - e \theta B\sp{tot}(\vec \eta, t)}, \\
\hat E_i\sp{tot}(\vec X,t) \, =\, & {E_i\sp{tot}(\vec \eta, t) \over 1 - e \theta B\sp{tot}(\vec \eta, t)}, \nonumber
\end{eqnarray}
with $\vec \eta$ defined by (3.1) and $F_{\mu\nu}\sp{tot}$
decomposing additively
\begin{equation}
F_{\mu\nu}\sp{tot} \, =\, F_{\mu\nu}\sp{CS}\, +\, F_{\mu\nu}\sp{ext}.
\end{equation}

As an obvious consequence of this procedure we see that
the minimal substitution (5.1) for the total gauge field defined by (5.3)
leaves the
symplectic structure (2.7) unchanged.

\subsection{Hybrid coupling}
In this case we couple the CS and external fields differently,
introducing
$A^{ext}_\mu$ %
into
the symplectic form as in (2.10). We assume
\begin{equation}
L\,=\,\tilde L\sp{CS}_{ext}\, +\, e(\hat A_i\sp{ext}\dot X_i\, +\, \hat A_0\sp{ext}),
\end{equation}
where $\tilde L\sp{CS}_{ext}$ is given by (2.20) with $\hat A_{\mu}$
replaced by $\hat A_{\mu}\sp{CS}$. In this coupling scheme, which
we call hybrid, the CS field is coupled via the minimal
substitution rule (2.19) while the electromagnetic background
field is coupled like in the Duval-Horvathy model.

If we consider the case of constant external fields $\hat
B\sp{ext}$ and $\hat E\sp{ext}$ we find that the second term in
(5.7) becomes, modulo a gauge dependent total time-derivative term,
\begin{equation}
e\,\left(\frac{1}{2}\hat B\sp{ext}\,\epsilon_{ij}\,X_i\,\dot X_j\,+\,\hat E_i
\sp{ext}\cdot X_i\right).
\end{equation}

Note that from the two terms in (5.8) the first one is known to be
invariant with respect to the time-independent area preserving
coordinate transformations ([12], [26]), but the second is not
invariant. However, we can further modify the action by adding the
following term proportional to $\theta$:
\begin{equation}
-\frac{e\sp2\theta}{2}\,\epsilon\sp{\mu\nu\rho}\,\hat F_{\nu\rho}\sp{ext}\,\hat A_{\mu}
\sp{CS}.
\end{equation}

With 
such a
term
we obtain instead of (5.8)
\begin{equation}
\frac{e\hat B\sp{ext}}{2}(\epsilon_{ij}
X_i\dot X_j \,  -\, 2e\theta \hat A_0\sp{CS})
+\, e \hat E_i\sp{ext}(X_i\,  +\, e\theta\,\epsilon_{ij}\hat A_j\sp{CS}),
\end{equation}
and we see that in the second term of (5.8)  $X_i$ has become replaced by the invariant
coordinate $X_i\, +\, e\theta\,\epsilon_{ij}\hat A_j\sp{CS}=\eta_i
(\vec X,t)$.

Note that (5.10) is quasi-invariant with respect to time-dependent area-preserving
transformations (2.21-22) 
\begin{equation}
\delta (\epsilon_{ij}X_i\dot X_j \,  -\, 2e\theta \hat A_0\sp{CS})
\,=\,e\theta\,\frac{d}{dt}\,(X_i\partial_i \Lambda\,-\,2\Lambda).
\end{equation}
So we have
\begin{equation}
L_{hyb}\,=\,\tilde L\sp{CS}_{ext}\,+\,\frac{e\hat B\sp{ext}}{2}(\epsilon_{ij}
X_i\dot X_j \,  -\, 2e\theta \hat A_0\sp{CS})\, +\, e\hat E_i\sp{ext}\cdot \eta_i.
\end{equation}

We would like to make the following comments:
\begin{enumerate}
\item[(i)] The additional terms (5.10) lead to the change of the symplectic structure
from (2.7) to (2.18) with $B=\hat B\sp{ext}$.

\item[(ii)] Expression (5.9) looks like the interaction of an induced current
\begin{equation}
J_{\theta}\sp{\mu}:\,=\,-\frac{e\theta}{2}\,\epsilon\sp{\mu\nu\rho}\,\hat F_{\nu \rho}\sp{ext}
\end{equation}
with the CS-gauge potential $\hat A\sp{CS}_{\mu}$. Obviously, the current
$J_{\theta}\sp{\mu}$ is conserved.

\item[(iii)] Arbitrary time-dependence of $\hat E_i\sp{ext}$  preserves the quasi-invariance of $L_{hyb}$ with respect to the transformations (2.21-22).
However, any space-dependence of $\hat F\sp{ext}_{\mu\nu}$ or time-dependence of $\hat B\sp{ext}$ spoils it.
\end{enumerate}

One can consider $L_{hyb}$ given by (5.12) for the critical value of the $B$
field i.e. at
\begin{equation}
\hat B\sp{ext}_{crit} = (e\theta)\sp{-1}.
\end{equation}
Then,
\begin{itemize}
\item
the two terms being proportional to $\hat A_0\sp{CS}$ 
in (5.12) add up to zero and so, due to the Gauss
constraint,
the $\hat A_i\sp{CS}$ becomes trivial, i.e. the CS field decouples from our
particles.
\item By the point transformation [10]
\begin{equation}
X_i\,\rightarrow\, q_i:\,=\, X_i\,+\, \theta \,\epsilon_{ik}\,P_k
\end{equation}
one finds as derived by Duval et al [10] that
\begin{equation}
L_{hyb}\,=\,\frac{1}{2\theta}\,\epsilon_{ij}\,q_i\dot q_j
\end{equation}
i.e. the particle phase-space reduces to two degrees of freedom. Furthermore, the particle EOM reduce to the Hall constraint [10]
\begin{equation}
P_i = e\theta \epsilon_{ij} E_j.
\end{equation}

We see, therefore, that in the critical case (5.14), even in the presence
of a CS-coupling, the Hilbert space 
reduces to the well known subspace
of the lowest Landau level describing the Quantum Hall Effect.
\end{itemize}

\section{Outlook}
\setcounter{equation}{0}

The aim of this paper has been to discuss the couplings with a gauge field
of our planar particle model [1,9] which provides,
via canonical quantization, noncommutative position coordinates
(see (2.7)). The relations (2.7) are
invariant under time-dependent area-preserving transformations (2.21).

In our  paper we have presented a coupling of Abelian gauge
fields which transform under generalized gauge transformations
(see (2.24)) in a way which implies the invariance of the action under
the joint transformations (2.21) and (2.22).
We have shown that after changing the phase space variables for point
planar particles and introducing classical SW transformation for
gauge fields one can identify our model with the one containing
gauge coupling as presented by Duval and Horvathy [9,10].
We would like to stress here that our classical SW transformation
(see (3.10-12)) relates the gauge fields formulated on
two noncommutative coordinate spaces (see (2.18) and (2.7))
which, only to the first order in $\theta$, coincides
with the standard SW transformations.

Our results on the two-body problem, with the inclusion of an external
magnetic field, should be further extended. Detailed quantum mechanical calculations along the lines given in a recent
paper by Correa et al [27] are called for.

The considerations presented in this paper describe nonrelativistic
dynamics in $2+1$ dimensions. In such a case the action (1.5) is
Galilean-invariant. The analogous relativistic model can
be constructed in $D = 1+1$.
 In a general $D$-dimensional relativistic case we could introduce
 the following extension of the action for a relativistic massless particle
\begin{equation}
{\cal L} = \frac{1}{e} \dot{X}^2_\mu - \frac{k}{e^2} \dot{X}_\mu
\ddot{X}_\nu \theta^{\mu\nu},
\end{equation}
where $\dot{X}_\mu \equiv \frac{dX_\mu}{ds}$ and 
 $s$ describes a  parametrization of the particle trajectory
and $e$ is an einbein variable transforming under reparametrization 
$s^\prime = s^\prime(s)$ by the formula
$e^\prime(s^\prime) = \left( \frac{ds^\prime}{ds}\right)^{-1}
 e(s)$. Unfortunately, if $\theta^{\mu\nu}$ is a constant,
 the action (6.1) breaks the $D$-dimensonal Lorentz
 invariance\footnote{We would like to mention that the relativistic
 invariance can be restored if we promote the constant $\theta^{\mu\nu}$
 to a one-dimensional field $\theta^{\mu\nu}(s)$ (see e.g. [28]).
 }.

One of the questions which should be also addressed is the second
quantization of the model (1.5), i.e. the passage from the classical
and quantum
mechanics to the corresponding field-theoretic model.

The required $D = 2+1$-dimensional field-theoretic model should have
the following properties\footnote{Such a model would help to solve
 the problem of the relation between the second Galilean  central
 charge and spin, recently discussed by Hagen [29]. A first attempt to construct such a model has been done very recently by Horvathy et al. [30].}:

i) In the limit $\theta \to 0$ it should become the Schr\"odinger
theory for free nonrelativistic $D=2$ particles.

ii) For $\theta \not= 0$ it should be invariant under the Galilei
group with two central charges, $m$ and $\theta$, and should lead to
the nonvanishing value of $\theta$ from the commutator of generators
of Galilei boosts.

Finally we would like to observe that  in this paper
we have dealt only with the couplings of Abelian gauge fields. In order
to consider coupled non-Abelian gauge fields  we would have to extend our model from [1]
by supplementing the space-time geometry by new degrees of freedom 
describing non Abelian charge space coordinates (see [31-34]).

\renewcommand{\theequation}{A.\arabic{equation}}
\annexe{1}{Appendix - Gauge Field Dependent Dreibein Formalism}


\par


\setcounter{equation}{0}
In this appendix we would like to derive
a gauge field-dependent dreibein formalism.

We  solve (\ref{luz3.3}) for ${\cal P}_k$ and so get
\begin{equation}\label{luz3.4}
  {\cal P}_k = \dot{X}_i E_{ik} + E_{0k}\, ,
\end{equation}
where
\begin{eqnarray}\label{luz3.5a}
  E_{ik} &=& ( 1 +  e\theta \widehat{B})^{-1}
(\delta_{ik} +        e\theta \varepsilon_{kj} \partial_i
\widehat{A}_{j}),
  \\
E_{0k} & = & e\theta \varepsilon_{ij} \partial_j \widehat{A}_{0}
E_{ik} \label{luz3.5b}
\end{eqnarray}
describes a dreibein differing from the one proposed
in [13], in the case of components
(\ref{luz3.5a}), only by an invariant factor. The dreibein components (\ref{luz3.5a}-\ref{luz3.5b})
transform with respect to the transformations (\ref{luz2.19}-22) as
follows:
\begin{equation}\label{luz3.6}
  \delta E_{\mu k} = e\theta \varepsilon_{ij} (\partial_\mu
  \partial_j \Lambda )E_{ik}\, ,
\end{equation}
which is a special case of the general 
transformation formula for a generic field $f(\overrightarrow{X},t)$ [18]

\begin{equation}\label{luz3.7}
\delta(\partial_\mu f) = \partial_\mu \delta f + e\theta
\varepsilon_{kj}(\partial_\mu \partial_j \Lambda )
\partial_k f \, .
\end{equation}

The formulae (\ref{luz3.5a}-\ref{luz3.5b}) can be treated as the
modification, with nonvanishing torsion, of the torsion-less
$\theta$-dependent dreibein presented in [13] (see [13], formula
(20)), with  the components $E_{00} =1$ and $E_{k0} = 0$ kept
unchanged.

The inverse dreibeins $e^\rho_\mu E_{\rho}^{  \ \nu} =
\delta_{\mu}^{ \ \nu}$ have a simple form
 ($ e_{\mu}^{  \ k} \equiv e_{\mu k}$)
\begin{eqnarray}\label{luz3.9}
  e_{0}^{\ 0} = 1\, , \qquad  e_{i}^{\ 0} = 0
  \cr
  e_{\mu k} = \delta_{\mu k} + e\theta \varepsilon_{ik}
  \partial_i \widehat{A}_\mu
\end{eqnarray}
and provide the formula for the derivative
\begin{equation}\label{luz3.10}
  D_\mu = e_\mu ^{\ \nu} \partial_\nu \, ,
\end{equation}
which is invariant under the local transformations (\ref{luz2.19}-22).

\end{document}